# Strongly Enhanced Sensitivity in Planar Microwave Sensors Based on Metamaterial coupling

Mohammad Abdolrazzaghi, *Student Member, IEEE*, Mojgan Daneshmand, *Senior Member, IEEE*, and Ashwin K. Iyer, *Senior Member, IEEE*

*Abstract*— Limited sensitivity and sensing range are arguably the greatest challenges in microwave sensor design. Recent attempts to improve these properties have relied on metamaterial- (MTM-) inspired open-loop resonators (OLRs) coupled to transmission lines (TLs). Although the strongly resonant properties of the OLR sensitively reflect small changes in the environment through a shift in its resonance frequency, the resulting sensitivities remain ultimately limited by the level of coupling between the OLR and the TL. This work introduces a novel solution to this problem that employs negative-refractive-index TL (NRI-TL) MTMs to substantially improve this coupling so as to fully exploit its resonant properties. A MTM-infused planar microwave sensor is designed for operation at 2.5 GHz, and is shown to exhibit a significant improvement in sensitivity and linearity. A rigorous signal-flow analysis (SFA) of the sensor is proposed and shown to provide a fully analytical description of all salient features of both the conventional and MTM-infused sensors. Full-wave simulations confirm the analytical predictions, and all data demonstrate excellent agreement with measurements of a fabricated prototype. The proposed device is shown to be especially useful in the characterization of commonly-available high-permittivity liquids as well as in sensitively distinguishing concentrations of ethanol/methanol in water.

*Index Terms*—artificial materials, metamaterials, resonators, couplers, metamaterial transmission-lines, signal-flow analysis, sensors, permittivity measurement

## I. Introduction

Microwave-based sensors have demonstrated enticing functionalities in many applications such as chemical, agricultural, medical, oil, and microfluidic systems [1-6]. They are attractive for their low cost, CMOS compatibility, easy fabrication, and design flexibility compared to other types of optical, thermal, cavity-based, and MEMS-based sensors. The real-time response of microwave sensors is a primary reason for their superiority over expensive and labor-intensive chemical procedures for *in situ* applications [7, 8].

In dielectric-constant measurements of pure/composite aqueous solutions, resonant methods [9], [10] are preferred to broadband methods [11], [12] because of their high accuracy. Several microwave resonator-based sensors, whose operation is principally based on dielectric-constant variation in the material under test (MUT), have been reported. For example, a probe tip loaded with a dielectric resonator has been used to detect concentrations of sodium chloride [13] and glucose [14] in deionized water. A microwave biosensor based on a cavity resonator has been developed for measuring pig blood D-glucose [15]. In addition, relatively high concentrations ($1\mu L$/ml) of biomolecules (*e.g.* streptavidin) were measured using planar biosensors [16], which also prompted DNA detection using the same process [17].

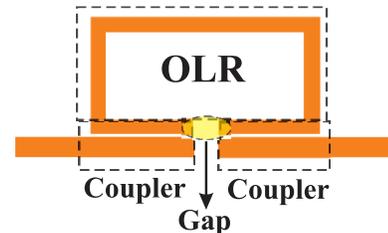

Fig. 1. Schematic of conventional planar resonator sensor depicted as cascade of couplers and an open-loop resonator (OLR) separated by a gap.

A relatively recent development in planar microwave-sensor design is the use of resonant structures known as open-loop resonators (OLRs) [18] or their modified configurations as split-ring resonators (SRRs) [19]–[23]. These elements, which were initially deployed as constituent elements of electromagnetic metamaterials (MTMs), are compact, exhibit high quality factors, and offer non-contact, robust sensing suited to harsh environments and applications involving small analyte volumes. These attributes have made them increasingly popular as transducers in various sensing applications [24]. As an added advantage, manipulation of the resonators' spatial configuration due to their flexible planar feature offers more convenience in practical setups compared with bulky and rigid waveguides/cavities. Many approaches, including that employed in this work, focus on the planar implementation of these resonators, in which they are typically edge-coupled to a

This work is supported financially by Alberta Innovates and Technology Futures (AITF), Natural Sciences and Engineering Research Council of Canada (NSERC), Canadian Microsystems Corporation (CMC) and Canada Research Chair (CRC).

Mohammad Abdolrazzaghi is with Electrical and Computer Engineering department of University of Alberta. 11-203 Donadeo Innovation Centre for Engineering, 9211-116 Street NW, Edmonton, Alberta, Canada (mabdolra@ualberta.ca)

Dr. Mojgan Daneshmand is currently an Associate Professor and Canada Research Chair Tier II Professor in the Dept. of Electrical and Computer Engineering, 11-324 Donadeo Innovation Centre for Engineering, University of Alberta, 9211-116 Street NW, Edmonton, Alberta, Canada (daneshmand@ualberta.ca)

Dr. Ashwin K. Iyer is currently an Associate Professor in the Dept. of Electrical and Computer Engineering, 11-211, Donadeo Innovation Centre for Engineering, University of Alberta, 9211-116 Street NW, Edmonton, Alberta, Canada (iyer@ece.ualberta.ca)

microstrip (MS) transmission line (TL), as shown in Fig. 1 [25]. In this arrangement, the strongly resonant properties of the OLR sensitively reflect small changes in the environment through a shift in its resonance frequency. However, the resulting sensitivities and sensing ranges are ultimately limited by the degree of coupling between the arms of the OLR and the adjacent TL, which affords only as much coupling as a typical microstrip coupled-line coupler, which in turn, can be simply modeled by a coupling capacitor in most planar-resonator circuit models [26], [27]. This equivalent coupling capacitance $C_c$ is well known to have a significant effect on the resonance frequency of coupled system [27], [28]. In fact, $C_c$ is also directly impacted by the MUT's permittivity; however, this effect is not prominent due to the typically low coupling of conventional coupled line couplers. The optimum conventional coupling length of $\lambda/4$ is too long to work very well for $\lambda/2$ resonators due to size and layout restrictions, and shorter lengths offer typically weaker coupling. Ideally, we require short couplers providing a high degree coupling. Although this is not achievable using conventional coupled-line couplers, high coupling become entirely feasible with MTM-based couplers.

Recently, negative-refractive-index TL (NRI-TL) MTMs [29], in which a host TL is periodically loaded with lumped capacitors and inductors to achieve a backward-wave response, have been used to realize novel couplers that exhibit nearly unity coupling with steeper phase-lead over extremely short lengths by exploiting the continuous contradirectional leakage of power between an NRI-TL and MS-TL [30]. Inspired by the NRI-TL coupled-line coupler, we introduce a novel planar microwave-sensor implementation in which the input MS-TL is replaced by a NRI-TL and edge-coupled to the long arm of the OLR. The resulting strong, contradirectional leakage of power is shown to dramatically enhance $C_c$, which in turn, imbues the planar sensor with enhanced sensitivity and linearity.

The design of such sensors and their responses often begins with full-wave simulations. This work employs the finite-element-method simulator Ansys HFSS [31]. However, as all constituent components of the proposed resonator are based on TL elements, we also propose a novel analytical technique employing microwave network analysis and signal-flow analysis (SFA), which is inspired by approaches previously employed to describe optical ring resonators [24], [26 - 27]. This approach is inherently suited to the microwave regime and is particularly elegant and powerful in predicting the response of the proposed MTM-infused sensors, where the constituent MTM couplers and lumped elements can be easily represented through their scattering or transfer matrices. Using the developed SFA, we are able to predict large enhancements in sensitivity as well as dramatically improved dynamic range for large permittivity variations. The SFA is closely validated by both full-wave simulations and measurements. Furthermore, it is shown that the proposed MTM-based sensor is superior to its conventional MS-TL counterpart in discriminating lossy MUTs - a feature that results from its improved dynamic range.

Section II establishes the theory of operation and design process for both the conventional and MTM-infused sensors, which are validated using a SFA and corroborated using full-wave simulations. Section III presents experimental results in two sub-sections; the first presents the proposed sensor's calibration fitting function for liquid-sensing applications while the second demonstrates the sensor's efficacy and sensitivity in detecting concentrations of methanol/ethanol in deionized water. Section IV concludes the paper.

## II. THEORY AND DESIGN

In this section, we establish analytically the operation of the conventional sensor architecture by viewing it as a microwave-coupler problem as described above. Thereafter, we explain how substituting conventional microwave couplers with NRI-TL MTM-based couplers, which offer much higher coupling over shorter lengths, enables dramatic enhancements in sensitivity ($\Delta f / \Delta \varepsilon_r$) and dynamic range (maximum detectable range of permittivity variation).

For the purposes of analysis, we interpret the sensor shown in Fig. 1 to consist of a combination of three components: an OLR, a gap, and two planar microstrip couplers. The core OLR is a microstrip half-wavelength resonator, which is bent into a rectangular shape to reduce its size. The small gap separating two open ends of the resonator establishes a strong capacitance at resonance. The coupling between the input/output TLs and the resonator is provided by the coupled-line couplers, which include the arms of the OLR. The resonance frequency of the sensor depends on the effective permittivity of the surrounding medium. Therefore, knowing *a priori* the nominal (i.e. vacuum) frequency response of the sensor, it becomes possible to determine the permittivity of the MUT. This determination may be made accurately provided that the quality factor of the resonator is sufficiently high. Commercially available full-wave electromagnetic simulation software used in the analysis of these types of structures can require extremely large meshes, particularly for finely-featured structures such as those involving MTMs, resulting in very long simulation times. To address this issue, we propose an analytical approach based on a SFA, which is fast, simple, and accurate in yielding the sensor's frequency response.

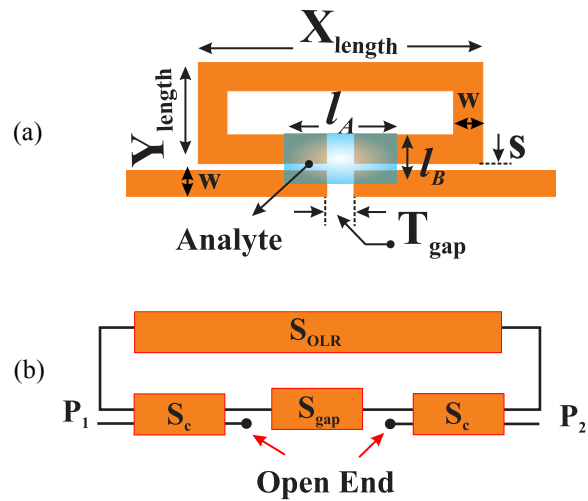

Fig. 2 (a) Dimensions of conventional sensor, (b) Representation as a cascade of 2- and 4-port networks.

Table I
Dimensions of the sensor

| VARIABLE | $X_{length}$ | $Y_{length}$ | $T_{gap}$ | $w$ | $s$ | $l_B$ | $l_A$ |
|---|---|---|---|---|---|---|---|
| Dim. (mm) | 20 | 7 | 1 | 2.4 | 0.4 | 4 | 10 |

*A. Signal-Flow Analysis*

*Schematic* - A half-wavelength resonator consisting of TLs with $Z_0 = 50\ \Omega$ is established on Rogers 5880 substrate ($\varepsilon_r = 2.2, \tan(\delta) = 0.0009, h = 0.8\ mm$) and the dimensions are given in Table I. The effective permittivity of the microstrip substrate is computed with accuracy better than 0.2% [35] and includes dispersive characteristics of the TL allowing the possibility of higher-order-harmonic generation [36]–[38]. In addition, the contributions of various sources of loss are included [39]–[41].

Fig. 2(a) presents the proposed sensor as an OLR that is edge-coupled to a split TL, a design that enables the possibility of a peak in the transmission spectrum at resonance. Each of the open ends of the split TL resembles an element with voltage reflection coefficient $\Gamma = 1$. The power is inserted from the left and is coupled to the OLR over a coupling region, and then circulates inside the OLR and gap, before the power is coupled again to the output TL and collected at the output.

A comprehensive and accurate modeling of the sensor requires modeling each constituent component in the network. Here, we employ an S-parameter representation:

*MS-TL/MS-TL coupler* – With port references as indicated in Fig. 3, the S-matrix of a conventional coupled-line coupler may be written as follows [40]:

$$S_{MS-MS} = \begin{bmatrix} 0 & S_{12} & S_{13} & 0 \\ S_{21} & 0 & 0 & S_{24} \\ S_{31} & 0 & 0 & S_{34} \\ 0 & S_{42} & S_{43} & 0 \end{bmatrix} \quad (1)$$

where

$$S_{21} = S_{12} = S_{43} = S_{34} = -je^{\frac{-j(\beta_e + \beta_o)l}{2}} \sin(\frac{(\beta_e - \beta_o)l}{2}) \quad (2)$$

$$S_{13} = S_{31} = S_{24} = S_{42} = e^{\frac{-j(\beta_e + \beta_o)l}{2}} \cos(\frac{(\beta_e - \beta_o)l}{2}) \quad (3)$$

Matching for all ports ($S_{11} = S_{22} = S_{33} = S_{44} = 0$) as well as perfect isolation ($S_{23} = S_{32} = S_{14} = S_{41} = 0$) are assumed in this analysis; these may, of course, be practically achieved through proper selection of the MS-TL dimensions, which control the even- and odd-mode impedances [40].

*OLR Components* - Representation of the input and output coupling through a four-port coupler description has already accounted for a portion of the path through the OLR. The remaining portion is considered as an isolated TL of length $l$ with propagation constant $\beta$ and S-parameters as follows:

$$S_{OLR} = \begin{bmatrix} 0 & Te^{j\beta l} \\ Te^{j\beta l} & 0 \end{bmatrix}, \quad (4)$$

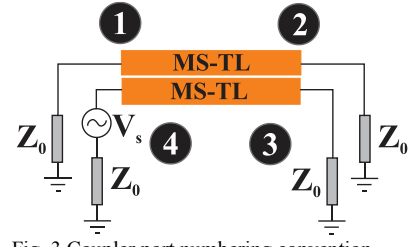

Fig. 3 Coupler port numbering convention.

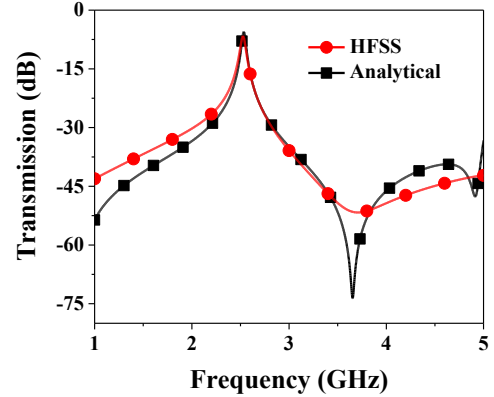

Fig. 4 Broadband fitting comparison between proposed SFA and HFSS.

where $T$ stands for the attenuation in the TL that accounts mainly for the conductor and dielectric losses as $T = e^{-\alpha_{total} l}$, where $\alpha_{total} = \alpha_c + \alpha_d\ (dB/cm)$.

The gap in the resonator is separately considered as a series capacitance $C_{gap}$ in a TL with S-parameters as follows:

$$S_{Gap} = \begin{bmatrix} \dfrac{Z}{2Z_0 + Z} & \dfrac{2Z_0}{2Z_0 + Z} \\ \dfrac{2Z_0}{2Z_0 + Z} & \dfrac{Z}{2Z_0 + Z} \end{bmatrix}, \text{ where } \begin{cases} Z = \dfrac{1}{j\omega C_{Gap}} \\ Z_0 = 50 \end{cases} \quad (5)$$

For the conventional microstrip sensor, a comparison between the proposed analytical method and full-wave simulation using HFSS shows a very good consistency (see Fig. 4), revealing both resonance peaks and antiresonances, over a wide range of frequencies from 1-5 GHz. Minor discrepancies at higher frequencies may be attributed to several simplifying assumptions, such as that of TEM mode propagation and the absence of higher-order TL modes. Also not considered for simplicity are parasitics introduced through microstrip bends, although equivalent lumped networks for these types of discontinuities may easily be included.

*B. Sensitivity Analysis*

This section investigates the effect on the sensor's transmission spectrum of a MUT introduced into the OLR gap. The MUT is modeled as an additional capacitance proportional to its permittivity, and the resulting downshifts in the resonance are validated using HFSS.

*Using SFA* - For the purpose of the SFA, the gap is modeled as a capacitance $C_{gap}$, which consists of the parallel combination of the microstrip gap capacitance ($C_0$) and a capacitance ($C_{var}$)

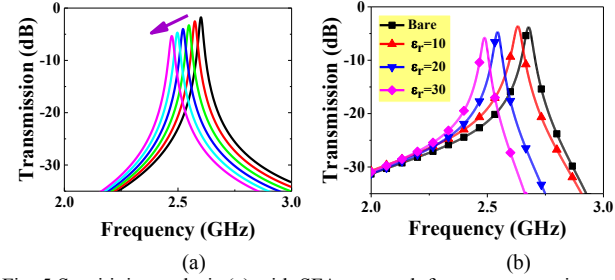

Fig. 5 Sensitivity analysis (a) with SFA approach for a representative variation in OLR's gap capacitance, and (b) permittivity variation in MUT simulated in HFSS

that is contributed by the MUT. Combination of the components using the SFA (in MATLAB) successfully predicted a resonant peak at 2.56 GHz for the bare sensor. A controlled sweep is done over the gap capacitance, which may be expressed as $C_{gap} = C_0 + C_{var}$, where $C_{var}$ is proportional to the MUT, permittivity, and $C_0$ is the capacitance of the OLR's bare gap. This results in the shifts shown in resonances in Fig. 5(a). With the increase in $C_{var}$, both the resonance frequency and the quality factor of the resonance are reduced, indicating the coupling between TLs and OLR, which causes more power to be trapped in the resonator than released to appear at the output [39].

*Using HFSS* – The performance of the sensor with respect to its sensitivity to the MUT is examined using an analyte with size $l_A \times l_B \times h_A$ ($l_A$=10, $l_B$=4, $h_A$=2 [mm]) on the OLR's gap, as shown in Fig. 2(a). This location offers the highest sensor sensitivity given its very high electric-field strength. A sweep over a selection of complex MUT permittivities ($\varepsilon_r = 1$ (Bare), 10, 20, 30, $tan(\delta) = 0.01$) is simulated in HFSS (Fig. 5(b)), where the assumed loss tangent is typical of realistic MUTs (higher losses are examined later in the work). As a result of increased capacitive loading due to the MUTs, the resonant spectrum undergoes a downshift, which matches the trend predicted by the SFA. This result shows that SFA captures the general behavior of the sensor with less complexity than time-consuming numerical methods.

Having established that the degree of (capacitive) coupling of the coupler determines the resonance downshift, a further simplification of the SFA (with associated reduction in accuracy) is to model the coupled-line coupler as a capacitor $C_c$ [41], [42], followed by a TL segment in parallel with the gap capacitance $C_{gap}$. This simple interpretation enables us to use a transmission- (ABCD- or T-) matrix formalism to determine the overall transfer function, as follows:

$$T_{total} = T_{C_c} \times T_{OLR+gap} \times T_{C_c} = \begin{bmatrix} A & B \\ C & D \end{bmatrix} \quad (6)$$

where

$$T_{C_c} = \begin{bmatrix} 1 & \frac{1}{j\omega C_c} \\ 0 & 1 \end{bmatrix}, T_{OLR} = \begin{bmatrix} \cos(\beta l) & jZ_0\sin(\beta l) \\ jY_0\sin(\beta l) & \cos(\beta l) \end{bmatrix} \quad (7)$$

and:

$$T_{OLR+gap} = \begin{bmatrix} -1 & j\omega C_{gap} + \frac{1}{jZ_0 \sin(\beta l)} \\ \frac{(2j\omega C_{gap}+1+\cos(\beta l))(1-\cos(\beta l))}{jZ_0 \sin(\beta l)} & 1 \end{bmatrix} = \begin{bmatrix} -1 & B' \\ C' & 1 \end{bmatrix} \quad (8)$$

The corresponding transmission parameter $S_{21}$ is therefore given by the following expression:

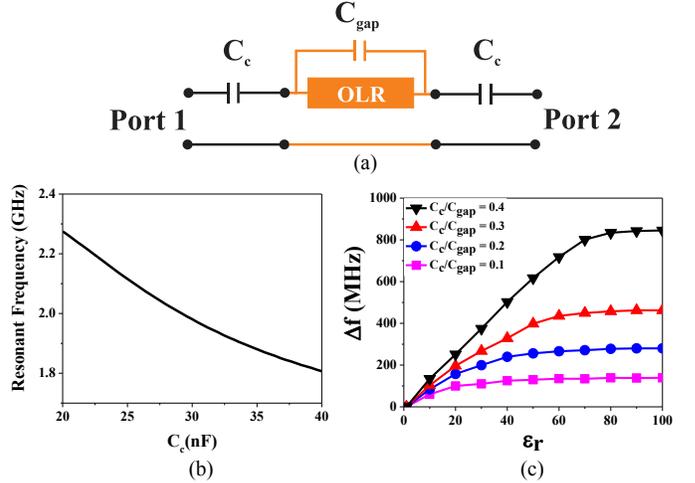

Fig. 6 (a) Coupling capacitors $C_c$ enabling 2-port resonator model (b) effect of $C_c$ on resonance frequency (c) effect of MUT on resonance frequency with respect to $C_c/C_{gap}$

$$S_{21} = 20 \log \left| \frac{2}{\frac{2C'}{j\omega C_c} + \frac{B'}{Z_0} + \frac{C'}{Z_0(j\omega C_c)^2} + C'Z_0} \right| \quad (9)$$

The resonance frequency ($f_{res}$) is defined as the local maximum of the transmission profile, which can be analytically derived by evaluating $\partial S_{21}/\partial f = 0$. As shown in the denominator of (9), $C_c$ has a significant role in determining $f_{res}$ of Fig. 6(a), which is shown with a typical variation, in Fig. 6(b). The shift in $f_{res}$ is an expected behavior due to $C_c$, and is pronounced for large capacitive coupling, which suggests a distinct benefit to improving this coupling mechanism. The effect of a MUT placed at the OLR gap may now be considered analytically using two additional capacitances parallel with both $C_{gap}$ and $C_c$. The shifts in resonance frequency for MUT permittivity values covering the range $\varepsilon_r = 1:100$ are presented in Fig. 6(c), where the effect of the coupled-line coupler is incorporated in the capacitance ratio $C_c/C_{gap}$ in the range of 0.1-0.4. The simulation for different coupling-capacitance ratios shows not only that larger $C_c$ yields larger variation in resonance frequency for a given variation in MUT permittivity, but also that the sensitivity and dynamic range is improved considerably for larger $C_c$.

The above result affirms the value of seeking methods of improving the coupling represented by $C_c$, and therefore by the coupled-line coupler. Here, we may invoke the NRI-TL/MS-TL coupler [30], which is known precisely for its nearly 0-dB contradirectional coupling over electrically very short lengths.

*NRI-TL/MS-TL Coupler* – The NRI-TL coupler geometry is constructed by placing an NRI-TL adjacent to a MS-TL, as described in [5], for which the unit cell is depicted schematically in Fig. 7(a). The NRI-TL portion consists of a MS-TL periodically loaded using series capacitors *C* and shunt inductors *L*. The isolated dispersion characteristics of representative NRI-TL and MS-TL designs are indicated in Fig. 7(b). The backward-wave behavior of the NRI-TL is evident from the negative slope of its dispersion curve (e.g., in the region from approximately 1 GHz to 3.5 GHz), in contrast to the forward-wave (conventional) dispersion of the MS-TL. Contradirectional coupling occurs where the two TLs are phase-matched, i.e. where their isolated dispersion curves intersect. The resulting coupled dispersion curves

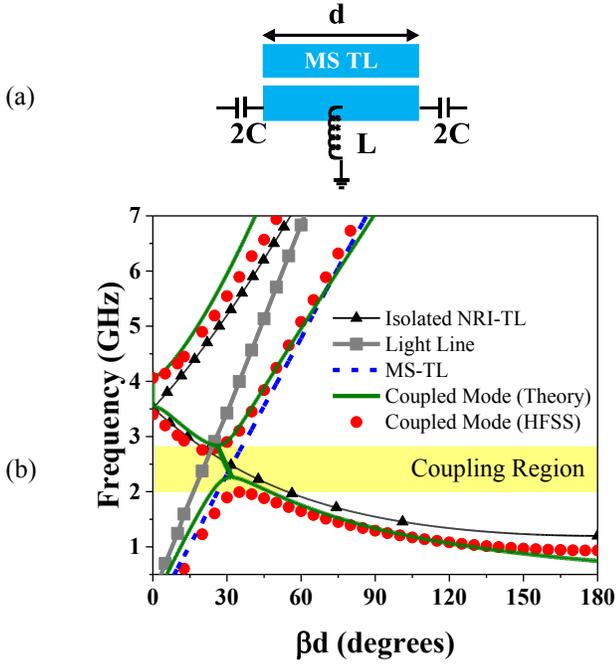

Fig. 7 (a) Equivalent circuit of NRI-TL MTM coupler; (b) Isolated- and coupled-mode dispersion diagrams.

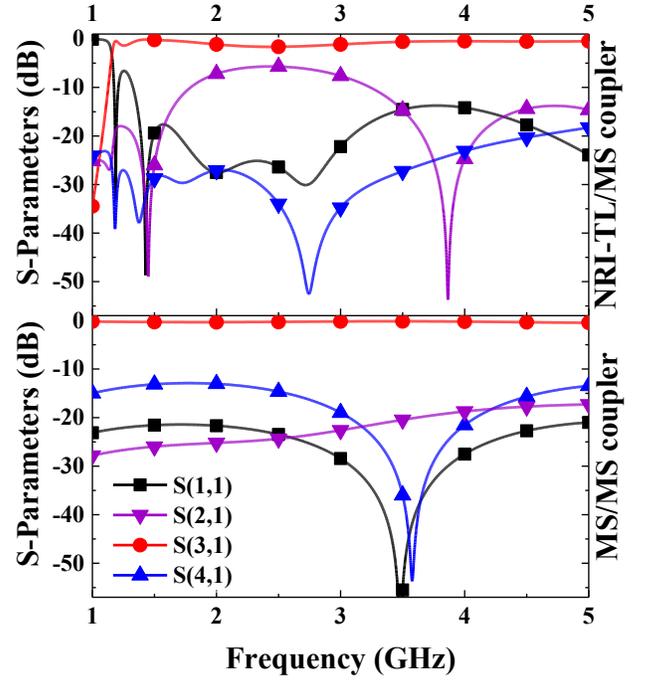

Fig. 8 S-parameters of NRI-TL/MS and MS/MS couplers (simulated in HFSS)

(computed analytically using a transmission-matrix formalism for an infinite periodic structure and also using HFSS) exhibit a complex-mode region, describing the leakage of power from the forward MS-TL mode to the backward NRI-TL mode, which is manifested as a bandgap. As a result, the coupled port of the NRI-TL/MS-TL coupler is adjacent to the input port.

Wideband impedance matching of the NRI-TL is possible using a design condition (also known as a "balanced" condition) for which the Bloch impedance of the periodic structure is matched to that of the underlying TL, which additionally ensures the stopband between the backward-wave and forward-wave passbands is closed [43].

In order to find the values for the loading elements $L$ and $C$, which impact the frequency range and dispersion of the complex-mode region, we should be mindful that the frequency range of interest be centered about the intersection of the MS-TL and NRI-TL dispersion curves; i.e., that

$$\beta d_{MS} = \beta d_{NRI-TL} \quad (10)$$

at $f_{res}$. As a starting point, we select $f_{res}$ = 2.5 GHz. The loading elements $L$ and $C$ may be found with respect to the band-gap closure condition for a given value of $d$ (length of unit-cell) using design guidelines outlined in [30]. Loading elements satisfying the above-noted $f_{res}$ and of values that may be easily obtained using off-the-shelf, surface-mount components are determined to be $C = 1.1 pF$ and $L = 2.73 nH$.

The S-parameters of the coupler in the complex-mode band with propagation constant $\gamma = \alpha + j\beta$ can be expressed as follows [30]:

$$S_{NRI-TL/MS} = \begin{bmatrix} 0 & S_{12} & 0 & S_{14} \\ S_{21} & 0 & S_{23} & 0 \\ 0 & S_{32} & 0 & S_{34} \\ S_{41} & 0 & S_{43} & 0 \end{bmatrix} \quad (11)$$

$$S_{12} = S_{21} = \frac{e^{+j\beta Nd}}{\cosh(\alpha Nd) + j \sinh(\alpha Nd)\cot(\varphi)} \quad (12)$$

$$S_{14} = S_{41} = S_{23} = S_{32} = \frac{1}{\cos(\varphi) - j \sin(\varphi)\coth(\alpha Nd)} \quad (13)$$

$$S_{34} = S_{43} = \frac{e^{-j\beta Nd}}{\cosh(\alpha Nd) + j \sinh(\alpha Nd)\cot(\varphi)} \quad (14)$$

where the NRI-TL is assumed to have $N$ unit cells, each of length $d$, and $\varphi$ is the transverse voltage phase difference between the two TLs for the c-mode and π-mode. Matched conditions at all ports ($S_{11} = S_{22} = S_{33} = S_{44} = 0$) and perfect isolation ($S_{13} = S_{31} = S_{24} = S_{42} = 0$) are assumed, and may be straightforwardly designed as outlined in [30].

The coupling performance of the NRI-TL/MS-TL is compared with that of the conventional MS-TL/MS-TL over total length of 40mm. Fig. 8 illustrates a considerably higher (backward) coupling of -4.5 dB at 2.5 GHz for the NRI-TL/MS-TL coupler (configured akin to that shown in Fig. 4(b)), while the MS-TL/MS-TL coupler only provides -12 dB (of forward coupling) at the same frequency.

Having established the superiority of the NRI-TL/MS-TL coupler in providing a high coupling level, this coupler may now be integrated into the sensor architecture. The embedded NRI-TL/MS-TL coupler in the coupling stage of the sensor (Fig. 2(a)) is designed with the dimensions given in Table II for an impedance level of $Z_0 = 50$ Ω, such that the resonance frequency would be slightly higher than the intersection point in Fig. 7(b), in order to allow downshifted resonances to reside entirely inside the complex-mode region. The loading capacitors $C$ are connected in series with microstrip TLs, and inductors $L$ are supported by narrow strips terminating in vias

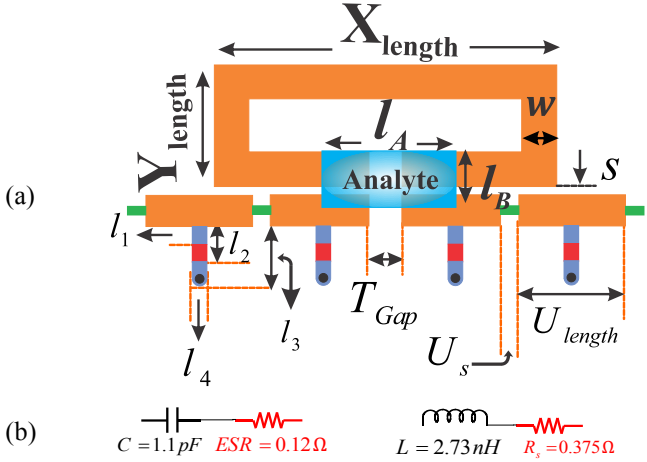

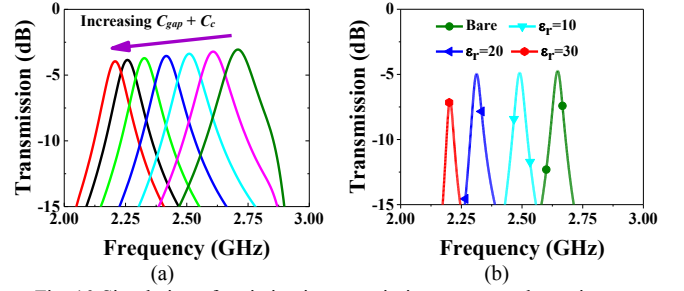

Fig. 10 Simulation of variation in transmission spectrum due to increase in (a) $C_{gap} + C_c$ with SFA (b) permittivity of MUT in HFSS.

Fig. 9(a) Proposed NRI-TL MTM-based sensor's dimensions (b) Loading-element models including loss.

Table II
Dimensions of the sensor

| VARIABLE | $X_{length}$ | $Y_{length}$ | $U_s$ | $w$ | $s$ | $T_{gap}$ | $l_A$ |
|---|---|---|---|---|---|---|---|
| Dim. (mm) | 20 | 7 | 1 | 2.4 | 0.4 | 1 | 10 |
| VARIABLE | $U_{length}$ | $l_B$ | $l_1$ | $l_2$ | $l_3$ | $l_4$ | $h_A$ |
| Dim.(mm) | 6.3 | 4 | 1 | 2 | 3 | 1.4 | 2 |

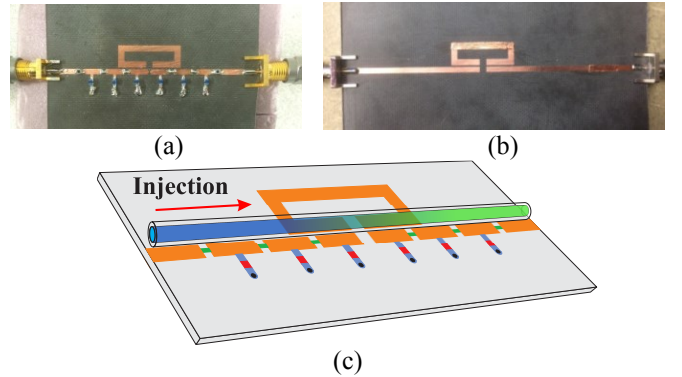

Fig. 11 (a) Fabricated NRI-TL MTM-based sensor; (b) Conventional sensor; (c) MUT-sensing setup.

(Fig. 9 (a)) to establish the shunt connection to ground. Sources of loss in the loading elements are considered according to the equivalent series resistance (ESR) of $C$ and quality factor of $L$ from datasheets, as shown in Fig. 9(b). It has already been confirmed that the operating frequency is well below the natural self-resonance frequencies of the elements selected for fabrication.

It has also been observed (although not shown here) that, whereas the conventional MS-TL/MS-TL coupler's coupling level varies with the MUT properties, the NRI-TL/MS-TL coupler retains a largely invariant coupling level, allowing it to operate predictably in a wider range of environmental conditions and applications.

*Sensitivity Analysis with SFA* – The SFA proposed earlier to describe the conventional sensor may be used with equal confidence for the NRI-TL/MS-TL-based sensor, given that the latter's scattering matrix (and therefore, its equivalent transmission matrix) is known. Permittivity variations are incorporated through specifying a $C_{var}$, which is placed in parallel with the gap capacitance $C_{gap}$. Fig. 10(a) shows the effect of incrementing $C_{var}$ on the transmission response of the sensor, where downshifts in $f_{res}$ are relatively large compared with their corresponding range for the conventional MS-TL/MS-TL sensor, shown in Fig. 5(a). This result confirms the higher sensitivity of the MTM-based sensor as a result of higher coupling afforded by the NRI-TL/MS-TL configuration.

*Sensitivity Analysis with HFSS* - The sensitivity performance of the sensor is examined by applying a MUT (analyte) as shown in Fig. 9(a). An analyte with volume $l_A \times l_B \times h_A$ ($l_A$=10, $l_B$=4, $h_A$=2 [mm]) is placed on the OLR's gap. A sweep over the MUT's dielectric and loss properties ($\varepsilon_r = 1, 10, 20, 30, tan(\delta) = 0.01$) is performed in HFSS (Fig. 10(b)). The resulting dramatic resonance shifts for the NRI-TL MTM-based sensor once again confirm the impact of enhanced coupling, which demonstrates nearly doubled sensitivity with respect to the conventional MS-TL/MS-TL, in qualitative agreement with the results shown for the SFA in Fig. 10(a).

In Sec. IIB, it was established that the four-port conventional coupled-line coupler could be reduced to an equivalent coupling capacitance $C_c$, and it was further shown that higher values of $C_c$ implied greater sensitivities. Given the strongly enhanced sensitivity observed for the MTM-based sensor, it is worthwhile to examine whether it can be associated with a commensurately high coupling capacitance. This analysis is conducted in the Appendix. A simplified expression for coupling is achieved by approximating the MTM coupler's properties using a two-port ABCD matrix approach. The ABCD analysis of the whole MTM-based sensor is then shown to agree very well with full-wave HFSS results, establishing the suitability of the MTM-based sensor for applications requiring high sensitivity. In the next section, the proposed NRI-TL/MS-TL sensor and the conventional MS-TL/MS-TL sensor are fabricated and their performance parameters are compared in a number of scenarios.

III. EXPERIMENTAL RESULTS

Both the NRI-TL/MS-TL and MS-TL/MS-TL sensors are fabricated on Rogers 5880 substrate with dimensions as listed in Table II, and in the former case, the capacitors are soldered in series and inductors are shorted using vias (Fig. 11(a-b)). Here, we examine the functionality of the sensors in two can be as high as 95%, mainly for MUTs with higher $tan(\delta)$. In order to have a robust calibration that can cope with a full range of materials, a mixed relation between these parameters is suggested as follows:

$$\varepsilon_r' = \sum_{i=0}^{M}\sum_{j=0}^{N} P_{ij} \left(\Delta f_{res}\right)^i \left(\Delta S_{21}\right)^j \quad (15)$$

$$\tan\delta = \sum_{i=0}^{M}\sum_{j=0}^{N} Q_{ij} \left(\Delta f_{res}\right)^i \left(\Delta S_{21}\right)^j \quad (16)$$

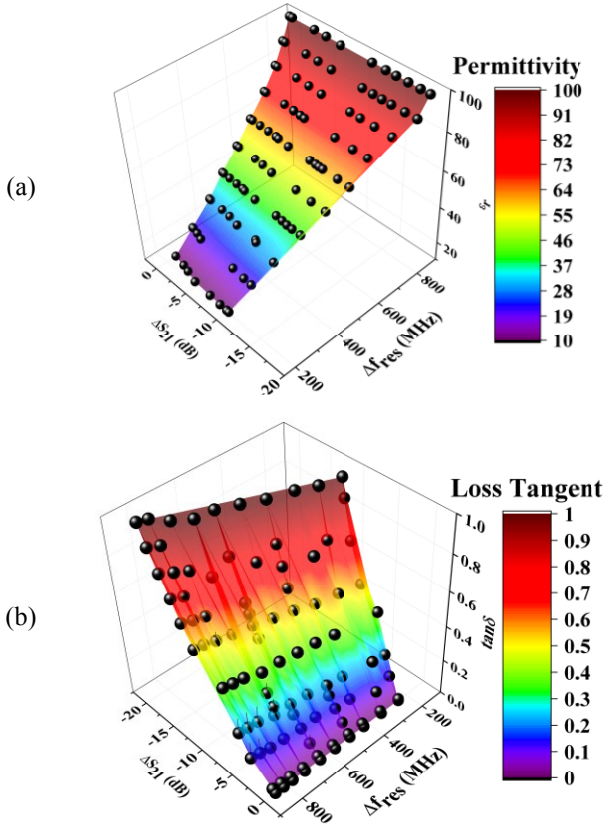

Table III
Coefficients of fitting functions for $\varepsilon_r'$ and $\tan(\delta)$

| Coef. | $P_{00}$ | $P_{01}$ | $P_{10}$ | $P_{20}$ | $P_{11}$ | $P_{02}$ | $P_{03}$ |
|---|---|---|---|---|---|---|---|
| Value | 53.81 | -1.45 | 26.86 | 4.81 | 0 | -0.31 | 0.074 |
| Coef | $P_{12}$ | $P_{21}$ | $P_{30}$ | | | | |
| Value | 0 | 0 | 1.57 | | | | |
| Coef | $Q_{00}$ | $Q_{01}$ | $Q_{10}$ | $Q_{20}$ | $Q_{11}$ | $Q_{02}$ | $Q_{03}$ |
| Value | 0.32 | -0.34 | -0.67 | 0.02 | 0.1 | 0.077 | -9e-3 |
| Coef | $Q_{12}$ | $Q_{21}$ | $Q_{30}$ | | | | |
| Value | -0.04 | -0.01 | -2e-3 | | | | |

where $P_{ij}, Q_{ij}$ are coefficients. A simplified version of this general type of polynomial fitting has been used in the literature [21] with M=N=1, but it is expanded up to M=N=3 such that $R^2 = 0.99$ can be obtained with the parameters listed in Table scenarios. First, the sensor is calibrated for monitoring liquids flowing inside a plastic tube placed along the coupling region (Fig. 11(c)). In the next section, the sensor is employed to detect extremely small concentrations of methanol/ethanol in deionized water in the same setup.

Fig. 12 Fitting surface for (a) permittivity and (b) loss tangent extraction from shifts in resonance frequency and amplitude of resonance.

*A. Measurement of Dielectric Properties*

In this section, simulation of the sensor using the proposed experimental configuration (material injection through a narrow tube), while varying the complex permittivity ($\varepsilon_r = \varepsilon_r' - j\varepsilon_r''$, $\tan(\delta) = \varepsilon_r''/\varepsilon_r'$) of the MUT, is conducted in HFSS. Generally, variations in the real part of the permittivity $\varepsilon_r'$ contribute to tuning of $f_{res}$, whereas the imaginary part $\Delta\varepsilon_r''$ primarily affects the magnitude of the transmission ($S_{21}$). However, in practice both the real and the imaginary parts of $\varepsilon_r$ contribute to $f_{res}$ and $S_{21}$ at $f_{res}$. The PTFE tube has inner and outer diameters of $\frac{1}{32}"\times\frac{1}{16}"$, respectively, and is placed along the OLR's gap (Fig. 11(c)). MUTs with $\varepsilon_r$ ranging from 1-100 and $tan(\delta)$ from 0-1 are passed through the tube, resulting in the variations in $\Delta f_{res}$ and $\Delta S_{21}$ that are depicted in Fig. 12(a) and (b).

It is worth examining the error introduced by considering only one of the two parameters ($\Delta f_{res}$ or $\Delta S_{21}$) in inferring $\varepsilon_r'$ and $tan(\delta)$. For example, it is determined that inferring $\varepsilon_r'$ using only $\Delta f_{res}$ (i.e., neglecting $\Delta S_{21}$) can lead to an error of up to 15%, mostly noticeable for lower $\varepsilon_r'$. Conversely, if only $\Delta S_{21}$ is used (i.e., neglecting $\Delta f_{res}$), the error in the loss tangent

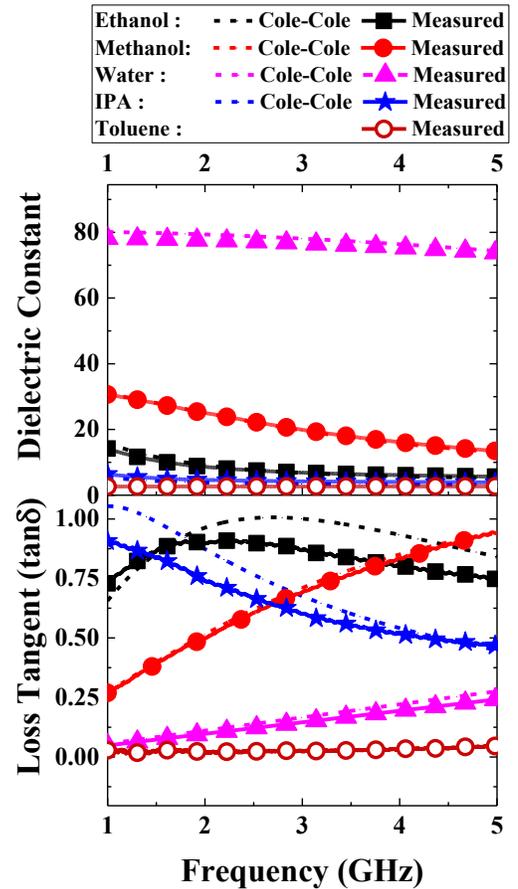

Fig. 13 Dielectric properties of materials using Cole-Cole equation and measurement using dielectric probe kit.

*Table IV*
Analytical and measured dielectric properties of known materials and comparing with the proposed fitting function

| Material | Cole-Cole | | Dielectric Probe | | MTM-Based Sensor | |
|---|---|---|---|---|---|---|
| | $\varepsilon_r'$ | $\tan\delta$ | $\varepsilon_r'$ | $\tan\delta$ | $\varepsilon_r'$ | $\tan\delta$ |
| Toluene | - | - | 2.5 | 0.02 | 2.41 | 0.025 |
| IPA | 4.38 | 0.76 | 3.56 | 0.67 | 4.1 | 0.71 |
| Ethanol | 7.56 | 1 | 7.69 | 0.9 | 7.71 | 0.96 |
| Methanol | 22.18 | 0.61 | 22.38 | 0.60 | 22.53 | 0.62 |
| Water | 77.19 | 0.139 | 78.8 | 0.123 | 76.7 | 0.134 |

(a) 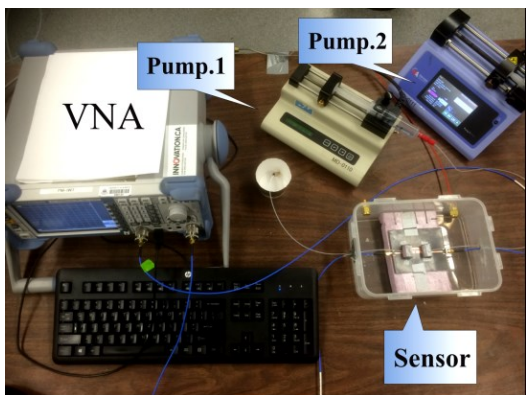

(b) 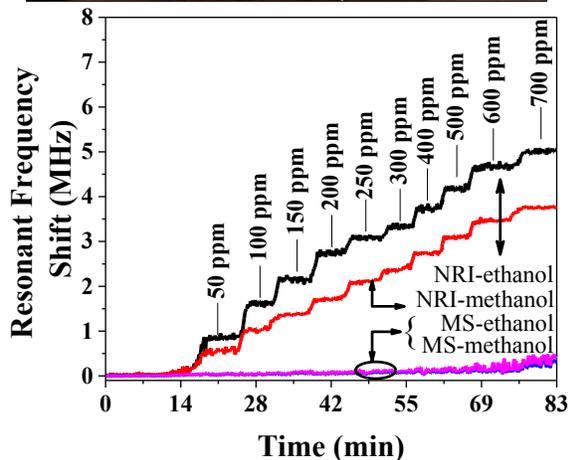

Fig. 14 (a) Experimental setup for concentration measruement; (b) Frequency shifts for various concentrations of ethanol/methanol in water.

III, and allows accurate identification of the MUT's dielectric properties.

To validate the MTM-based sensor's fitting function, we now compare the results of Fig. 12(a)-(b) with the known dielectric properties of some commonly available materials (IPA ($C_3H_8O$), ethanol($C_2H_6O$), methanol($CH_3OH$), water ($H_2O$) and toluene ($C_7H_8$)). These are obtained using the Cole-Cole equations [44], which we further corroborate with measurements obtained using an open-ended dielectric probe (Fig. 13) demonstrates the accuracy of the proposed sensor, and suggest that this approach will prove useful for the inference of the dielectric properties of a wide range of materials when employing the NRI-TL MTM-based sensor as well.

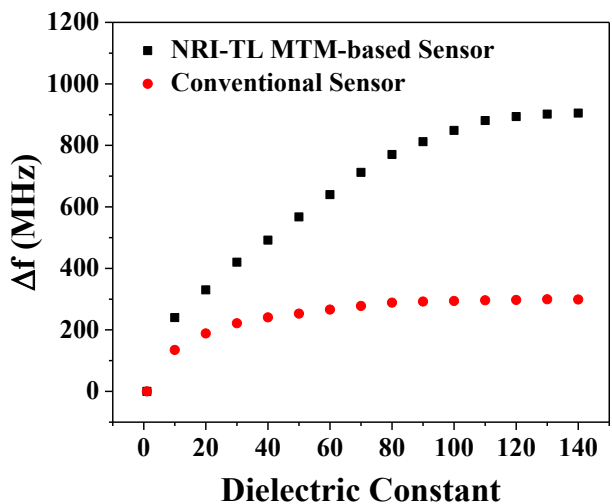

Fig. 15 Comparison of frequency shifts for both sensors with change in dielectric constant of MUT ($\tan\delta = 0$)

### B. Measurement of Ethanol/Methanol Concentrations

Fig. 14(a) presents another setup wherein the sensitivity and dynamic range of the sensor is demonstrated using minute samples of ethanol/methanol in deionized water administered according to the following process: two pumps are prepared to inject water and ethanol simultaneously into PTFE tubes that are joined together with a Y junction. Flow rates are designed to achieve the desired concentrations within the range of 50 – 700 ppm, while the mixture is passing over the gap of OLR (see Fig. 11(c)) and the waste goes to a paper mug bin. Dry air flow is used continuously in order to reduce the destructive effect of humidity in the experiment, such that a relative humidity of 0% is maintained throughout the experiment. The increase in ethanol concentration in the mixture shifts the frequency upwards; this may be expected from the fact that the increase in the volume fraction of ethanol (which has lower dielectric constant), and hence the effective permittivity of the mixture is reduced, accordingly increasing $f_{res}$. The transmission profiles are recorded in LabView connected to vector network analyzer (VNA). The resonance frequencies are extracted with methanol as the solvent over both of the sensors and the results are shown in Fig. 14(b). The differences between the data corresponding to the NRI-TL MTM-based sensor and those of its conventional counterpart are indeed dramatic, demonstrating considerable improvements in dynamic range and sensitivity. Moreover, it is evident that the MTM-based sensor can detect concentrations as low as 50ppm while the conventional sensor's performance saturates in the vicinity of the permittivity of water. The simulated resonance-frequency shifts, particularly for materials with larger permittivity (only the tan(δ)=0 case is shown for simplicity), are presented in Fig. 15 and demonstrate the proposed MTM-based sensor is especially suited to the sensitive discrimination of higher-permittivity materials.

The transmission magnitude and phase response of the sensors obtained from HFSS coincide with the analytical response predicted by the SFA and are shown to be in good agreement with measurement results, as shown in Fig. 16. This consistency shows the capability of the proposed SFA-based theory in predicting the full frequency response of the sensor quite accurately as a function of the dielectric properties of applied MUTs. Nevertheless, it is worth noting potential sources of discrepancy, such as the presence/propagation of quasi-TEM or higher-order modes in the microstrip structure, as well as radiation losses, ohmic losses and inherent tolerances in the loading components.

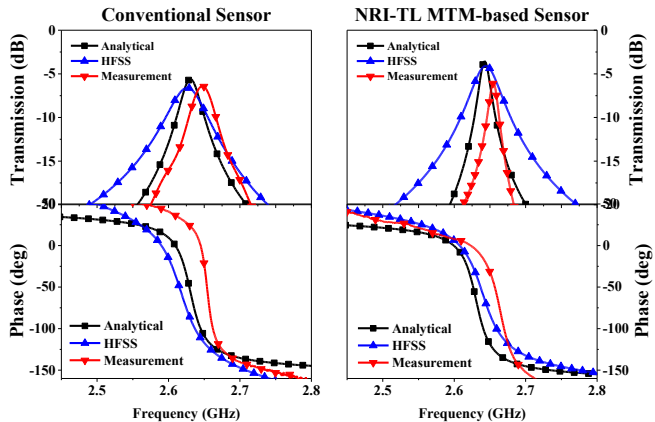

Fig. 16 Comparison of transmission magnitudes and phases between HFSS simulations and SFA with measurements of both sensors.

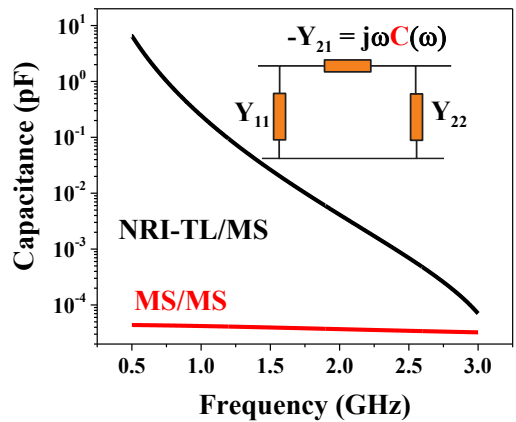

Fig. 17 Capacitance approximation from analytical examination of NRI-TL/MS-TL coupler

## IV. CONCLUSION

In this work, we proposed a novel planar microwave sensor based on NRI-TL MTMs coupled to OLRs. The NRI-TL MTM is dispersion-engineered to enable a wide frequency range supporting transmission resonances of the MUT-loaded OLR. An SFA is developed to predict the transmission response of the NRI-TL MTM-based sensor by cascading the S-parameters of each of the sensor's components. The proposed sensor is shown to exhibit very high sensitivity and dynamic range versus the conventional MS-TL sensor with respect to introduction of external MUTs exhibiting a wide range of complex permittivities.

The same analysis is done in HFSS and is confirmed by measurement results of transmission profile (magnitude and phase) for both sensors. The sensors are fabricated and tested with various commonly available and well-characterized chemicals as a means for calibration, and thereafter, fitting curves are extracted to be used for permittivity characterization for unknown MUTs. Concentration measurements of ethanol/methanol in water validated the highly superior performance of the proposed sensor in sensitively differentiating concentrations, particular for high-permittivity materials, in a water host medium. The proposed methodology for fabricating planar sensors can be applied universally to many other configurations and be used in applications requiring high sensitivity, such as blood-glucose monitoring or biomolecule detection.

## APPENDIX

In order to analyze the frequency dependence of the coupling levels of MS-TL/MS-TL and NRI-TL/MS-TL couplers when incorporated into the sensor, a two-port π-model of the coupling behavior (inset of Fig. 17) is studied. The coupling element of $-Y_{21}$ is found to be purely capacitive (Fig. 17). It is evident that the larger capacitance at lower frequencies demonstrated by the NRI-TL/MS-TL coupler is responsible for the downshift of resonant frequencies for MUTs with higher permittivity values, in contrast with the lower capacitance values for the MS-TL/MS-TL coupler, which also don't tend to vary much over frequency. This is an explicit advantage of MTM-based couplers over conventional ones due to their higher backward coupling levels.

In pursuit of a closed-form transmission response of the NRI-TL MTM-based sensor, an ABCD-matrix description of the circuit in Fig. 6(a) is employed. In this approach, the input coupler, OLR, and the output coupler are considered to be in series, instead of the more physically accurate interconnections

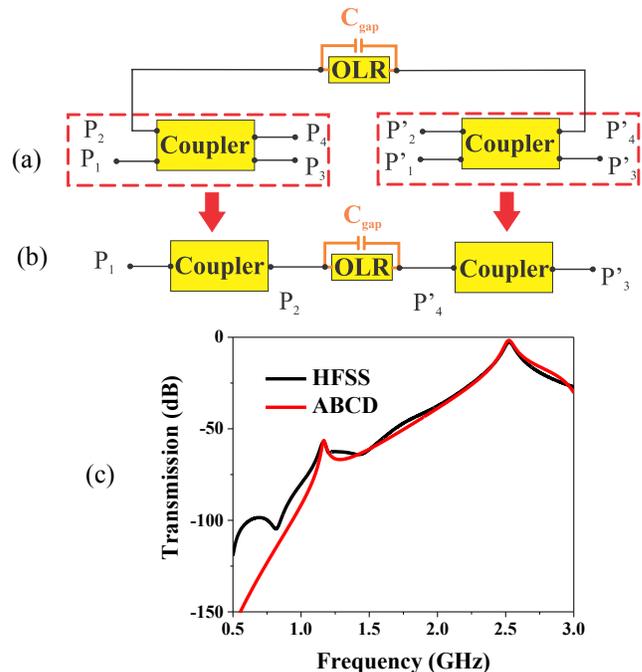

Fig. 18 (a) Sensor model employing 4-port couplers; (b) simplified model employing equivalent 2-port representation of couplers; (c) comparison of transmission response predicted by simplified model and HFSS simulations.

employed in the SFA. This approximation is validated by the fact that the high directivity of the MTM coupler yields a negligible leakage of power to $P_4$ in Fig. 18(a). Cascading the components in this manner prompts us to convert the four-port coupler into a two-port (Fig. 18(b)), after which the whole sensor's ABCD matrix may be easily calculated. Accordingly, $P_4$ (for high directivity of MTM coupler) and $P_3$ (split gap in TL) are terminated with open circuits, such that the ABCD matrix of the NRI-TL MTM-based coupler may be reduced from $T_{4\_port}$ (17) into $T_{2\_port}$ (18) as follows:

$$T_{4\_port} = \begin{bmatrix} a_{11} & a_{12} & a_{13} & a_{14} \\ a_{21} & a_{22} & a_{23} & a_{24} \\ a_{31} & a_{32} & a_{33} & a_{34} \\ a_{41} & a_{42} & a_{43} & a_{44} \end{bmatrix} \qquad (17)$$

$$T_{2\_port} = \begin{bmatrix} \dfrac{a_{22} - a_{32} \times \dfrac{a_{21}}{a_{31}}}{a_{12} - a_{11} \times \dfrac{a_{32}}{a_{31}}} & \dfrac{a_{22} - a_{12} \times \dfrac{a_{21}}{a_{11}}}{a_{32} - a_{12} \times \dfrac{a_{31}}{a_{11}}} \\ \dfrac{a_{42} - a_{12} \times \dfrac{a_{21}}{a_{11}}}{a_{12} - a_{11} \times \dfrac{a_{32}}{a_{31}}} & \dfrac{a_{42} - a_{41} \times \dfrac{a_{12}}{a_{11}}}{a_{32} - a_{12} \times \dfrac{a_{31}}{a_{11}}} \end{bmatrix} \qquad (18)$$

Thus, the total ABCD matrix is evaluated using the following relation:

$$T_{total} = T_{2-Port} \times T_{OLR+gap} \times T_{2-Port} = \begin{bmatrix} a & b \\ c & d \end{bmatrix} \qquad (19)$$

where the coupled-line coupler is modeled as $T_{2-port}$, and the transmission ($S_{21}$) is easily defined as follows:

$$S_{21} = 20 \log \left| \dfrac{2}{a + \dfrac{b}{Z_0} + cZ_0 + d} \right| \qquad (23)$$

The resonance frequency occurs where $S_{21}$ achieves a maximum; in other words, where the denominator of $S_{21}$ approaches zero. The resulting transmission profile demonstrates generally excellent agreement with HFSS simulations (for a representative case in Fig. 18(c)).

ACKNOWLEDGMENT

The authors would like to acknowledge Rogers Corporation, Coilcraft, and American Technical Ceramics (ATC) for free substrate, inductor, and capacitor samples, respectively. Moreover, many thanks to Dr. Justin Pollock for his constructive guidance in HFSS simulations.